\documentclass{emulateapj}
\usepackage{graphicx}
\begin{document} 

\title{The Observational and Theoretical Tidal Radii of Globular Clusters in M87}
\author{Jeremy J. Webb, Alison Sills, William E. Harris}
\affil{Department of Physics and Astronomy, McMaster University, Hamilton ON L8S 4M1, Canada}
\email{webbjj@mcmaster.ca}
\keywords{galaxies: individual (M87) - galaxies: kinematics and dynamics - globular clusters: general}

\begin{abstract}
Globular clusters have linear sizes (tidal radii) which theory tells us are determined by their masses and by the gravitational potential of their host galaxy. To explore the relationship between observed and expected radii, we utilize the globular cluster population of the Virgo giant M87. Unusually deep, high signal-to-noise images of M87 are used to measure the effective and limiting radii of approximately 2000 globular clusters. To compare with these observations, we simulate a globular cluster population that has the same characteristics as the observed M87 cluster population. Placing these simulated clusters in the well-studied tidal field of M87, the orbit of each cluster is solved and the theoretical tidal radius of each cluster is determined. We compare the predicted relationship between cluster size and projected galactocentric distance to observations. We find that for an isotropic distribution of cluster velocities, theoretical tidal radii are approximately equal to observed limiting radii for $R_{gc} < 10$ kpc. However, the isotropic simulation predicts a steep increase in cluster size at larger radii, which is not observed in large galaxies beyond the Milky Way. To minimize the discrepancy between theory and observations, we explore the effects of orbital anisotropy on cluster sizes, and suggest a possible orbital anisotropy profile for M87 which yields a better match between theory and observations. Finally, we suggest future studies which will establish a stronger link between theoretical tidal radii and observed radii.

\end{abstract}


\section{Introduction \label{Introduction}}

Historically it has been assumed that the gravitational field of the host galaxy regulates its satellite sizes, including globular clusters \citep[e.g.][]{vonhoerner57, king62, innanen83, jordan05, binney08, bertin08}.The size of a globular cluster is more commonly referred to as the \textit{tidal radius}, and has both a theoretical and observational definition. Theoretically, the tidal radius is better known as the Jacobi radius, which marks the distance past which, a star will feel a stronger acceleration towards the galaxy and escape the globular cluster. First-order tidal theory determines the tidal radius  \citep{vonhoerner57} via:

\begin{equation}\label{rtHoerner}
r_t=R_{gc}(\frac{M}{2M_g})^{1/3}
\end{equation}

\noindent where $R_{gc}$ is the galactocentric distance of the cluster, M is the the cluster's mass, and $M_g$ is the mass of the galaxy, assumed in early studies to be a point mass. For an isothermal halo, we have $M(R_{gc}) \propto R_{gc}$ and so to first order we should expect $r_t \propto R_{gc}^{\frac{2}{3}}$ if the mean cluster mass does not vary strongly with galactocentric distance. In addition, \textit{if} the structural properties of the clusters such as their central concentrations \textit{c} do not vary systematically with $R_{gc}$ either, then the mean effective (or half-mass) radius $r_h$ should also increase as $r_h \propto R_{gc}^{\frac{2}{3}}$. 

\begin{figure}[tbp]
\centering
\includegraphics[width=\columnwidth]{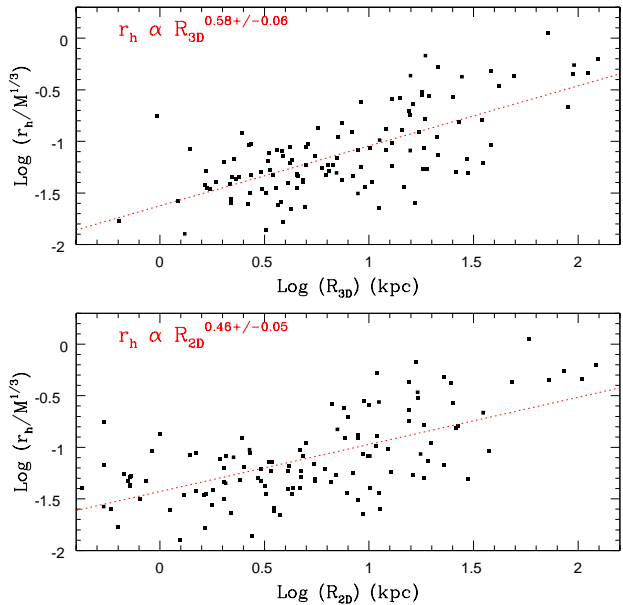}
\caption{$Log (r_h/M^{\frac{1}{3}})$ versus the log of each cluster's three dimensional galactocentric distance (top panel) and projected galactocentric distance (bottom panel). The red dashed lines mark the linear lines of best fit.}
  \label{fig:rh_norm}
\end{figure}

A rough illustration can be drawn directly from the Milky Way. Taking Milky Way cluster effective radii from Harris 1996 (2010 Edition), we plot the mass-normalized radius $log \ r_h/(M^{\frac{1}{3}})$ versus the log of each cluster's three dimensional galactocentric distance $R_{3D}$ (top panel) and projected galactocentric distance  $R_{2D}$ (bottom panel) in Figure \ref{fig:rh_norm}. We assume the Y-Z plane of the Milky Way is the plane of the sky. See \citet{vdbergh91} or \citet{mackey05} for earlier versions of this plot.  The purpose of normalizing cluster size by cluster mass is to remove any scatter due solely to differences in cluster mass as seen in Equation \ref{rtHoerner}. The lines of best fit indicate that $r_h \propto R_{3D}^{0.58 \pm 0.06}$ and $r_h \propto (R_{2D})^{0.46 \pm 0.05}$ for a given cluster mass. Although the projection to 2D does reduce the slope of $<r_h>$ to something close to $R^{\frac{1}{2}}$, it certainly does not flatten the trend entirely. While the relationship between cluster  $r_h$ and three dimensional galactocentric distance is comparable to the prediction from tidal theory, we do not expect an exact match as the Milky Way is not a spherical galaxy, and clusters do not have circular orbits as required in the calculation. The remaining scatter around the lines of best fit is expected to be due at least partly to differences in cluster orbits and central concentrations.

The bottom panel will be what we may expect to see for any giant galaxy that is projected onto the plane of the sky. In essence, basic tidal theory predicts that clusters in the outer halo of a giant galaxy are permitted to have much larger linear sizes than those in the inner halo, other things being equal. 

However, an intriguing puzzle beginning to emerge from recent measurements in a variety of giant E galaxies is that the observed trend is very much shallower than expected, near $r_h \propto R_{gc}^{0.1-0.2}$, where now we denote $R_{gc}$ as the projected (2D) galactocentric distance. \citep[e.g.][]{gomez07, harris09b}. See also \citet{spitler06} and \citet{harris10} for similar results from the giant Sa galaxy M104. These results do not yet have a full explanation.

These newly measured trends hint that other factors may be in play in addition to the simple tidal theory outlined above. In this paper, we explore particularly one possible route to explaining the observed shallow trend of $<r_h>$, which is to invoke an anisotropy gradient in the cluster orbits. \textit{If} outer halo clusters tend to have more strongly radial orbits than the inner-halo ones, they will be carried deep into the potential well of their parent galaxy and thus trimmed back to smaller-than-expected radii. More will be said about this in Section 3 below.

The observational size of a globular cluster is the \textit{limiting radius}, which is ``the outer limit of the cluster where the density drops to zero'' \citep{binney08}. For clusters outside of the Milky Way, the only way of defining the exact size of a cluster is to examine its density profile, or surface brightness profile. This is done by fitting observed profiles to models such as the well known \citet{king62} (K62), \citet{king66} (K66), \citet{wilson75} (W75), or \citet{sersic68} (S68) models.

While it is generally assumed that the Jacobi radius and the limiting radius of a cluster are the same, recent studies \citep[e.g][]{brosche99, gieles08, kupper10, baumgardt10} are finding that this assumption may require modification. A way to explore this assumption is to directly compare the relationship between cluster size and galactocentric distance for both an observed and simulated cluster population. If the correct relation can be established, then we can begin to utilize globular clusters in new ways to extract further information, such as the mass distribution within a galaxy and the orbital distribution of the clusters.

As an initial test case, we use the globular cluster population around the Virgo giant M87, which provides an exceptionally large number of clusters whose sizes are well resolved by the Hubble Space Telescope (HST). HST Archive images of M87 are available that are unusually deep and high signal-to-noise, containing nearly 2000 of its globular clusters. From a theoretical point of view the gravitational field, which is required for the calculation of a cluster's tidal radius, is better known for M87 than for any other giant elliptical galaxy \citep {mclaughlin99}. In addition, the foreground reddening and field contamination are low \citep{tamura06}. Essentially, M87 provides the best available testbed for comparing theoretical and observational tidal radii.

In Section 2 we will fit the observed surface brightness profiles of M87 globular clusters with different models, determine the tidal and effective radii of each cluster, and establish the observed trend between cluster size and galactocentric distance. In Section 3 we make use of observationally determined parameters of M87 to simulate a theoretical cluster population. Using the known gravitational field of M87, we can determine each simulated cluster's perigalactic distance and calculate theoretical tidal radii. Projecting the simulated clusters onto a two dimensional plane, we will then finally have a predicted relationship between cluster size and projected galactocentric distance. The results of the simulation are then compared to the observational results for both isotropic and anisotropic distributions. Our conclusions and future work are then discussed in Section 4.

In what follows we adopt $(m-M)_0 = 30.95$ for M87, to keep consistency with the mass distribution model of \citet{mclaughlin99}.

\section{Observations \label{stwo}}

The HST ACS/WFC Archive images we use in this study are from program GO-10543 (PI Baltz). The co-added composite exposures in each filter were the same ones described in detail in \citet{bird10}, constructed through use of the APSIS software \citep{blakeslee03}, which performs accurate image registration, cosmic-ray rejection, and distortion correction with drizzle. While the raw images are the same as those used by \citet{madrid09}, \citet{peng09} and \citet{waters09}, subpixel resampling was done during the drizzle step to yield final combined science images with an improved scale of 0\farcs 025~px$^{-1}$ (half the native pixel size of the camera). As will be seen below, the subsampling produced a noticeable improvement in the effective spatial resolution of the data compared with all previous studies \citep[see][for a more detailed description]{bird10}.

As an initial step, the ELLIPSE and BMODEL functions within STSDAS were used to fit elliptical isophotes to the brightness distribution of M87 in both the V and I images. Subtracting the isophotal model from the images to remove the brightness of the galaxy allowed for the easier detection of objects that were previously hidden (see Ferrarese et al. 2006 or Peng et al. 2009 for similar examples).

The residual images in F814W and F606W were then used to identify globular cluster candidates. After determining the standard deviation of the background sky pixel values, we searched for candidates brighter than threshold cut-offs chosen to be faint enough to include all true clusters but bright enough to exclude almost all individual halo stars in M87 itself. The \citet{bird10} analysis discusses the much more challenging measurement of the faint halo stars. After objects were matched based on their position on both images, we manually removed candidates very near the center of M87 ($R_{gc} < 0.2$ kpc or 100 pixels) and near the M87 jet, as the background light intensity is much higher in these regions. This resulted in 2052 globular cluster candidates. 

\begin{figure}[tbp]
\centering
\includegraphics[width=\columnwidth]{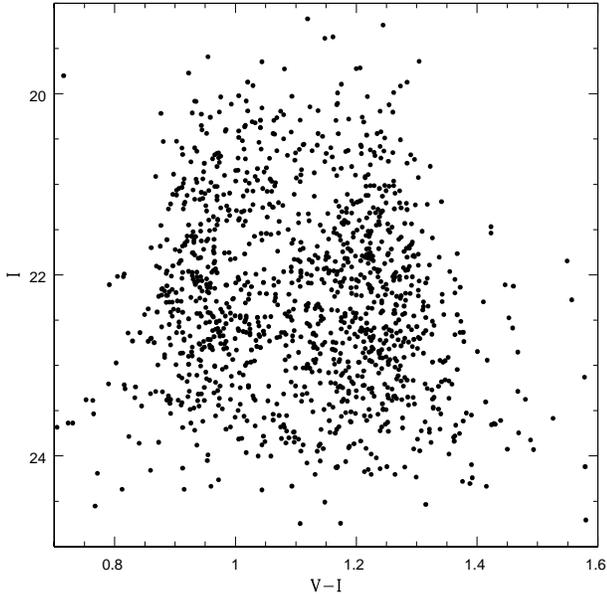}
\caption{CMD of the globular cluster candidates in M87. True magnitudes were obtained through aperture photometry extrapolated to large radius and converted to V and I. The familiar blue and red sequences are clearly visible.}
  \label{fig:HRdiagram}
\end{figure}

To construct a color-magnitude diagram (CMD) of the candidates, instrumental magnitudes were converted to true magnitudes through aperture photometry extrapolated to large radius \citep{sirianni05}, and converted to (V,I) with the transformations in \citet{saha11}. The CMD of the candidates is shown in Figure~\ref{fig:HRdiagram}. It reveals the familiar blue (metal-poor) and red (metal-rich) sequences that are now well established. For a more detailed analysis of the cluster photometry, we refer the reader to \citet{peng09}.

With the list of remaining globular cluster candidates, the surface brightness distribution of each cluster was fit with PSF-convolved K62, K66, W76, and S68 models via the code GRIDFIT  \citep[e.g.][]{mclaughlin08, harris10, barmby07}. The code returned best-fit values of effective radius ($r_h$) and central concentration which for the King models is the familiar $c=log(\frac{r_t}{r_c})$. To remove non-clusters we eliminated objects with $\chi^2$ values greater than 10 and all candidates with $c < 0.5$ and $c > 3.0$, which indicates a poor fit to the observations. Finally, we removed objects with large differences between the effective radius in the V and I bands, leaving us with 1290 globular clusters identified with high confidence.

We have carried out size measurements on both the original and isophote-subtracted images. Focusing on clusters in the outer regions of M87, where the light gradient is shallow and the influence of the subtraction is minimal, we found the mean offset ($r_h$(original)-$r_h$(subtracted)) to be 0.13 pc. Since the surface brightness profiles in the subtracted image are internally more precise, we take the effective radius as determined by model fits to clusters in the subtracted image and add the mean offset. This results in our cluster sizes being comparable to those of \citet{peng09} and \citet{madrid09}, with a mean difference of approximately 0.03 pc between both studies.

\begin{figure}[tbp]
\centering
\includegraphics[width=\columnwidth]{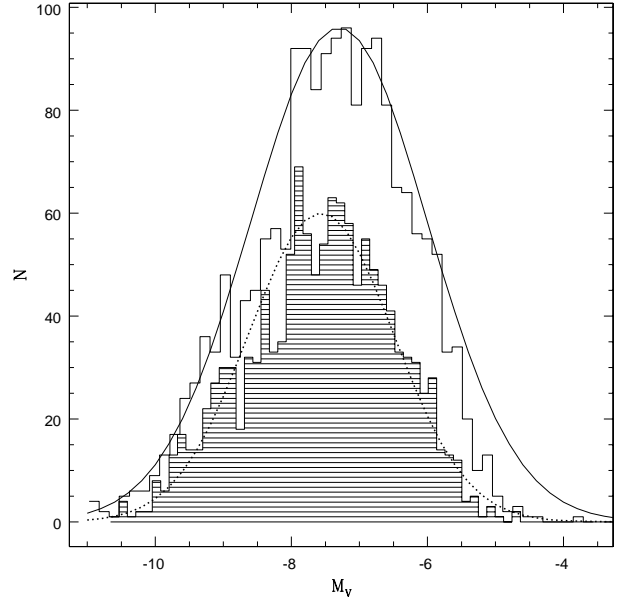}
\caption{V magnitude distribution of all the globular cluster candidates (un-shaded) and candidates which are not eliminated by the $\chi^2$, central concentration and $|\Delta r_h(V-I)|$ criteria (shaded). The usual globular cluster luminosity function with a turnover magnitude of $M_v = -7.3$ and a standard deviation of 1.3 is shown as a solid line. The luminosity function of the final candidate list is Gaussian-like, symmetric about a visual magnitude of $M_v = -7.6$ with a standard deviation of 1.0 (dotted line). }
  \label{fig:mv_hist}
\end{figure}

To examine which objects were removed from the globular cluster candidate list, we consider the V magnitude distribution of objects before and after the $\chi^2$, central concentration and $|\Delta r_h(V-I)|$ cuts were made. In Figure \ref{fig:mv_hist} we see that the original dataset (un-shaded) is in agreement with the usual globular cluster luminosity function, which has a turnover magnitude of $M_v = -7.3$ and a standard deviation of 1.3 \citep[e.g.][]{brodie06, peng09}. However, after the candidates have been removed based on the cuts described above, the distribution is reasonably approximated by a Gaussian-like function, symmetric about a visual magnitude of $M_v = -7.6$ with a standard deviation of 1.0 (dotted line), which we use below to set up our model simulation. It appears that model fits to the brightness profiles of faint clusters provide less accurate central concentrations or effective radii. We stress that this sub-selected luminosity function shape is used \textit{only} for the purposes of setting up our simulated globular cluster distribution.

\begin{figure}[tbp]
\centering
\includegraphics[width=\columnwidth]{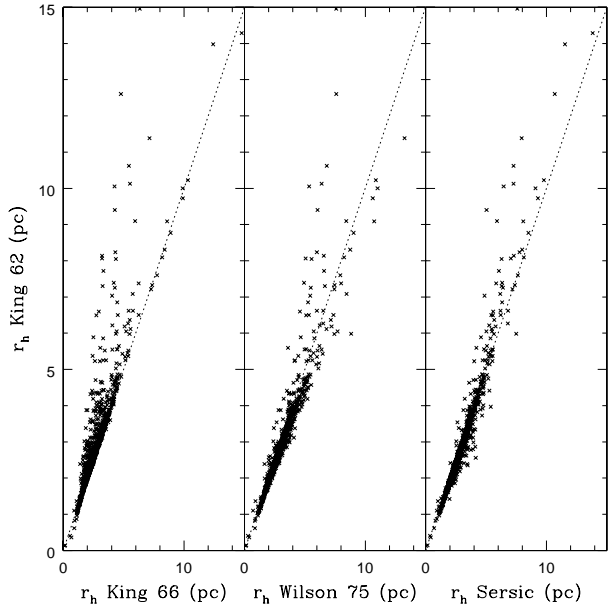}
\caption{K62 effective radius vs. K66 effective radius (left panel), W75 effective radius (center panel) and S68 effective radius (right panel). The dotted line represents equality.}
  \label{fig:rh_compare}
\end{figure}

Comparing the results of the K62, K66, W75, and S68 model fitting for the final list of globular cluster candidates (Figure \ref{fig:rh_compare}), we found that for many objects, the K66 tidal and effective radii were significantly smaller than those predicted by the other models. Inspecting the brightness profiles of these specific clusters, we observed that clusters with brightness profiles that are more extended were the source of the discrepancy. This is a known fault of K66 models, as they predict a sharp cutoff at the tidal radius while W75 models can more successfully fit a more gradual decrease in brightness or density near the tidal radius. Since K62 and S68 models are not dynamically motivated, they are  less affected by how cluster brightness behaves near the tidal radius. This led us to reject K66 model fits. Next, we inspected the mean and root mean square $|\Delta r_h(V-I)|$ for each cluster as found by all four models. K62 model fits yielded the lowest root mean square scatter, suggesting K62 fits have the best internal consistency. Therefore we adopted K62 measurements to compare with our simulations of M87 cluster sizes.

\begin{figure}[tbp]
\centering
\includegraphics[width=\columnwidth]{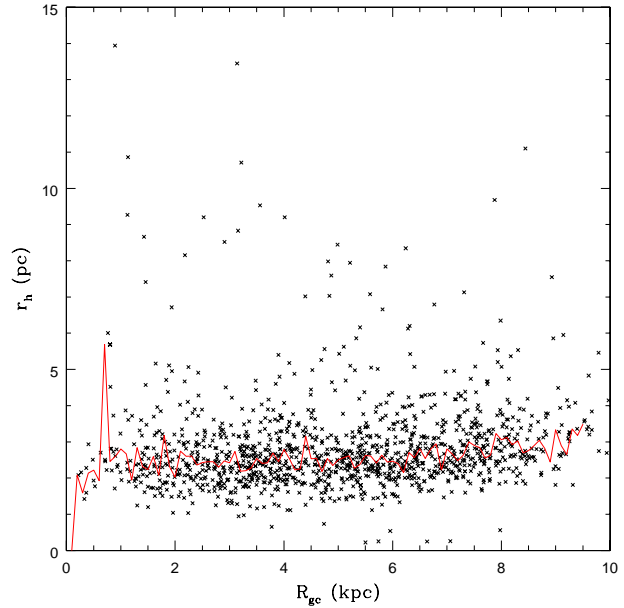}
\caption{K62 effective radius vs. log projected distance for observed globular clusters. The solid red line indicates the median effective radius calculated using radial bins that are 0.1 kpc in size.}
  \label{fig:rh_k62}
\end{figure}

The average K62 effective radius of each cluster in the V and I bands is plotted against projected galactocentric distance in Figure \ref{fig:rh_k62}. We choose to compare effective radii as opposed to tidal radii because effective radii $r_h$ are the directly measured quantity whereas limiting radii $r_t$ are only calculated from $r_h$ and c \citep[see][for further discussion]{harris10}. Looking at the relationship between effective radii and projected galactocentric distance, we find that the basic trend is much flatter than predicted by tidal theory or the one found for Milky Way clusters in Figure \ref{fig:rh_norm} (bottom panel).

\section{Simulation \label{sthree}}

\subsection{The Isotropic Case}

We now simulate a globular cluster population orbiting within the galactic potential of M87, in order to calculate theoretical radii as a function of $R_{gc}$. Each globular cluster was given a position in the halo ($R$, $\theta$, $\phi$), velocity ($v_r$, $v_\theta$, $v_\phi$), mass (M), and central concentration (c). The distribution parameters used in the simulation are drawn from Gaussians with parameters summarized in Table \ref{table:gcsim}. The spatial distribution was taken from \citet{harris09}, who found that the projected radial profile of the blue and red globular cluster subpopulations could be fit with a standard Hubble profile relating density ($\sigma_{cl}$) to projected distance ($R$):

\begin{equation}\label{hprof}
\sigma_{cl}(R)=\sigma_0 / (1+\frac{R}{R_0})^{-a}
\end{equation}

The appropriate values for $\sigma_0$,  $R_o$, and $a$ are listed in Table \ref{table:gcsim} for the blue and red subsystems. The angular distribution was assumed to be spherically symmetric.

The mass distribution of globular clusters was taken from the observed and culled globular cluster luminosity function (see Figure \ref{fig:mv_hist}). Assuming a mass-to-light ratio of 2 (e.g. \citet{mclaughlin05}), the resulting mass distribution is Gaussian about a mean of $log \frac{M}{M\odot} = 5.24$ and a standard deviation $\sigma (log \frac{M}{M\odot}) = 0.4$. In assigning a King-model central concentration parameter ($c = log \frac{r_t}{r_c}$) to each globular cluster, the distribution of central concentrations in the Milky Way from Harris 1996 (2010 Edition) was used, a Gaussian distribution with a mean of $\bar{c} = 1.5$ and standard deviation of 0.4. 

The velocity dispersion ($\sigma$) in spherical coordinates (R, $\theta$, $\phi$) is taken from the observed line of sight velocity dispersion of globular clusters in M87 \citep{cote01}. We have initially assumed the distribution of orbits to be isotropic, such that the anisotropy parameter ($\beta$) is zero and $\sigma_R = \sigma_\theta = \sigma_\phi$, where $\beta =1-\frac{\sigma_\theta^2+\sigma_\phi^2}{2 \sigma_R^2}$ \citep{binney08}. For comparison, the solved orbits of Milky Way globular clusters \citep{dinescu99, dinescu07} do not indicate a preference towards circular or radial orbits. In the later discussion, however, we relax this assumption and explore anisotropic distributions.

\begin{table}
  \caption{Simulated Globular Cluster Population Input Parameters}
  \label{table:gcsim}
  \begin{center}
    \begin{tabular}{lcc}
      \hline\hline
      {Parameter} & {Value} \\
      \hline

Radial Distribution & Hubble Profile \\
Blue Population & \\
$\sigma_0$ & 66 arcmin$^{-2}$ \\
$R_0$ & 2.0' \\
a & 1.8 \\
Red Population & \\
$\sigma_0$ & 150 arcmin$^{-2}$ \\
$R_0$ & 1.2' \\
a & 2.1 \\
Angular Distribution & Spherically Symmetric \\
Mass-To-Light Ratio & M/L = 2 \\
Mass Distribution & Gaussian \\
$\langle log(M/M_0) \rangle$ & 5.24 \\
$\sigma_{log(M/M_0)}$ & 0.40 \\
Velocity Dispersion & Gaussian \\
$\langle v \rangle$ & -19 km/s \\
$\sigma_v$ & 401 km/s \\
$\beta$ & 0 \\
Central Concentration & Gaussian \\
$\langle c \rangle$ & 1.5 \\
$\sigma_c$ & 0.4 \\  
      \hline\hline
    \end{tabular}
  \end{center}
\end{table}

Exactly 6000 globular clusters were simulated. Not only does this provide a statistically significant number of clusters, but also results in the same number of clusters within 10 kpc of M87 as the observed dataset. $40 \%$ were designated as ``red" clusters and had positions drawn from the red Hubble profile from Table \ref{table:gcsim}. The remaining $60 \%$ were designated as ``blue" clusters and were drawn from the appropriate observed parameters in Table \ref{table:gcsim}. The ratio of total number of blue clusters to red clusters is in agreement with the Hubble profiles in \citet{harris09}.

An isotropic distribution of cluster velocities has a broad range of orbital eccentricities (e.g. \citet{vdbosch99}). The normal approach to calculating tidal radii is to assume the tidal radius of a cluster is imposed at perigalacticon, where the tidal field of the host galaxy is the strongest. This assumption was initially suggested by \citet{vonhoerner57} and later \citet{king62}, and we will use it here. The assumption follows from the fact that for almost all real clusters, a cluster's relaxation time ($t_{r_h}$) is greater than its radial period, such that a cluster returns to perigalacticon before it is able to relax. We note, however, that recent studies \citep[e.g.][]{brosche99, kupper10} suggest that instead some sort of orbit averaged distance may result in a more accurate tidal radius.

Using the simulated parameters of each cluster, we calculate the theoretical tidal radius of each cluster as derived by \cite{bertin08}:
\begin{equation} \label{rt}
r_t=(\frac{GM}{\Omega^2\upsilon})^{1/3}
\end{equation}

\noindent Where $\Omega$, $\kappa$ and $\upsilon$ are defined as:

\begin{equation}
\Omega^2=(d\Phi_G(R)/dR)_{R_p}/R_p
\end{equation}
\begin{equation}
\kappa^2=3\Omega^2+(d^2\Phi_G(R)/dR^2)_{R_p}
\end{equation}
\begin{equation}
\upsilon=4-\kappa^2/\Omega^2
\end{equation}

\noindent $\Phi_G$ is the galactic potential due to the mass profile of M87, $R_p$ is the cluster's perigalactic distance, $\Omega$ is the orbital frequency of the cluster, $\kappa$ is the epicyclic frequency of the cluster at $R_p$, and $\upsilon$ is a positive dimensionless coefficient. The mass profile of M87 was taken from \citet{mclaughlin99}:

\begin{equation}\label{mt}
M_{total}(r)= M_{stars}(r) + M_{dark}(r)
\end{equation}
\begin{equation}\label{ms}
M_{stars}(r) = 8.10 \times 10^{11} \ M_{\odot} \ [\frac{(r/5.1kpc)}{(1+r/5.1kpc)}]^{1.67}
\end{equation}
\begin{equation} \label{md}
M_{dark}(r)= 7.06 \times 10^{14} \ M_{\odot} \times[\ln(1+r/560kpc)-\frac{(r/560kpc)}{(1+r/560kpc)}
\end{equation}

Since clusters are randomly assigned a position, mass, central concentration, and velocity, we must also determine which clusters are expected to survive to present day. More specifically, clusters with orbits that bring them to small perigalactic distances may either evaporate over a Hubble time due to tidal dissolution or be eliminated by dynamical friction. Each cluster's tidal dissolution time ($t_{dis}$) was calculated assuming $t_{dis} = 30 \times t_{rh}$, where $t_{rh}$ is a cluster's half-mass relaxation time determined with Equation 6 from \citet{meylan01}. The factor of 30 is consistent with NBODY simulations by \citet{trenti07} and within the range for a cluster in a tidal field from \citet{binney08}. Clusters with tidal dissolutions times less than 10 Gyr were rejected from the simulation, as they would have fully evaporated and not be observable today. Similarly, after calculating each cluster's infall time due to dynamical friction \citep{binney08}, we eliminate all clusters with infall times less than 10 Gyr.

\begin{figure}[tbp]
\centering
\includegraphics[width=\columnwidth]{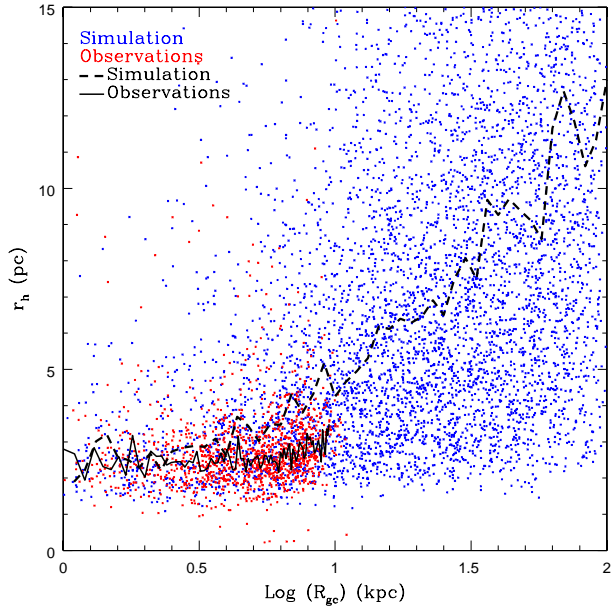}
\caption{Effective radius of each simulated globular cluster (blue) compared with its projected distance. The dashed black line indicates the median effective radius calculated using radial bins that increase in size at a rate of $10^{0.4}$ kpc. The observed clusters (red) and median (solid black line) from Figure \ref{fig:rh_k62} are also plotted.}
  \label{fig:rhsmooth_B0}
\end{figure}

To best compare to observations, we first convert all tidal radii to effective radii by assuming each simulated cluster can be represented by a K62 model. We then project the three dimensional position of each cluster onto a two dimensional plane, which represents the plane of the sky. Each simulated cluster's effective radius is plotted against its projected galactocentric distance in Figure \ref{fig:rhsmooth_B0}. For comparison purposes, we have also plotted each observed cluster in red and the observed median as a solid black line.

\subsection{Anisotropic Cases}

Up to this point, we have been operating under the assumption that the velocity distribution of M87 is isotropic ($\beta=0$). However, the value of $\beta$ may differ in different regions of a galaxy. For M87, \citet{cote01} found that while the cluster population of M87 appears to be isotropic as a whole, the inner regions of M87 could possibly have a negative value for $\beta$ such that orbits are preferentially tangential ($\beta < 0$), while the outer regions of M87 may contain clusters with more radial orbits ($\beta > 0$). \citet{weijmans09} made similar conclusions from observations of NGC 3379 and NGC 821. Theoretical work regarding the Milky Way by \citet{prieto08} and on dark matter haloes by \citet{zait08} and \citet{ludlow10} among others all find that $\beta$ increases from the isotropic case in the inner regions of a galaxy to more radial orbits in the outer regions. Additionally, many of these same studies all find evidence for $\beta$ to be less than zero at small galactocentric distances.

The simulation outlined above was repeated for different values of $\beta$. For $\beta$ greater than zero, the radial velocity dispersion was assumed to be equal to the observed velocity dispersion \citep{cote01}, and the $\theta$ and $\phi$ distributions were assumed to be equal ($\sigma_\theta = \sigma_\phi$) and calculated via the $\beta$ equation. Since $\sigma_R > \sigma_\theta$, clusters are brought deeper into the tidal field of the galaxy to smaller perigalactic distances. This in turn reduces their tidal and effective radii. For $\beta$ less than zero, the $\theta$ and $\phi$ velocity distributions are assumed to be equal to the observed velocity distribution, with the radial velocity distribution determined via the $\beta$ equation. Since $\sigma_\theta > \sigma_R$, clusters will have large tangential velocities and small radial velocities, keeping them far from the galactic center.

\section{Comparing Theory and Observations \label{sfour}}

\subsection{The Isotropic Case}

The first comparison that was made between observational and theoretical cluster radii was for the simulated isotropic case, with $\beta=0$. The median effective radius versus projected galactocentric distance is shown as the dashed black line in Figure \ref{fig:rhsmooth_B0}. Simulated clusters (blue), as well as the observed clusters (red) and observed median line (solid black line) are also plotted. The first observation that can be made is that the simulation and observations agree in the inner regions of M87. The second observation which can be made is that in the mid-halo to outer regions of M87, the simulation appears to overestimate observed cluster sizes. However, at larger galactocentric distances theoretical tidal radii may be larger than observed tidal radii, as it is possible that some clusters may be tidally under-filling. If they formed at a large galactocentric distances where the tidal field is weak, and it is likely that they will stay tidally under-filling for the duration of their lifetimes \citep{gieles10}. Due to the limited field of view of the observations, we cannot yet make comparisons past $R_{gc} \sim 9$ kpc. 

\subsection{Anisotropic Cases}

\begin{figure}[tbp]
\centering
\includegraphics[width=\columnwidth]{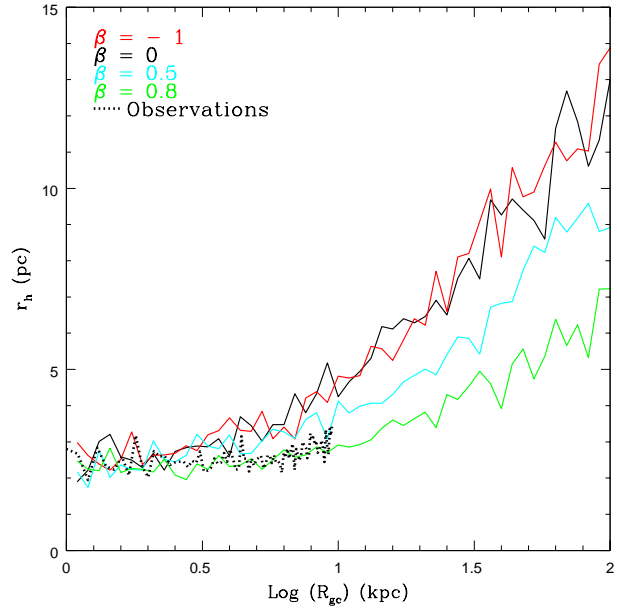}
\caption{Effective radius of each globular cluster compared with its projected distance for different values of $\beta$. The dotted black line is the median effective radius taken from K62 model fits to the observations (Figure \ref{fig:rh_k62}).}
  \label{fig:rhsmooth_Bcompare_wobs}
\end{figure}

We next compare the observational results to simulations with $-1 < \beta < 1$ as shown in Figure \ref{fig:rhsmooth_Bcompare_wobs} with the median observational K62 effective radii shown as a dotted black line. Along each of the model curves, $\beta$ is held constant with $R_{gc}$ at the value shown (that is, there is no $\beta$-gradient). Simulations with $\beta = -0.5$ and $\beta = 0.2$ were also performed. As expected, the $\beta=-0.5$ results are between the $\beta=-1$ and $\beta=0$ median lines in Figure \ref{fig:rhsmooth_Bcompare_wobs}, while the $\beta=0.2$ results are between $\beta=0$ and $\beta=0.5$.

Since none of the single $\beta$ simulations result in a perfect match between theory and observations, Figure \ref{fig:rhsmooth_Bcompare_wobs} suggests that an anisotropy profile $\beta(R_{gc})$ must be incorporated into the simulation. Matching simulated and observed effective radii suggests that the globular cluster population is approximately isotropic ($\beta = 0 $) for $R_{gc} < 2$ kpc, and radially anisotropic ($\beta > 0$) for $R_{gc} > 2 kpc$. To determine a possible anisotropy profile for M87, we perform a $\chi^2$ test between the observations and simulation in different radial bins. 

\begin{figure}[tbp]
\centering
\includegraphics[width=\columnwidth]{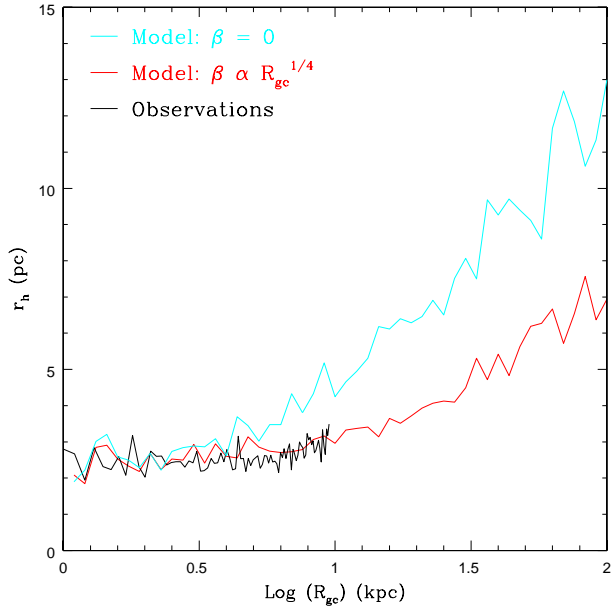}
\caption{Effective radius of each globular cluster compared with its projected distance for a radially dependent $\beta(r)$ (red) and for $\beta = 0$ (cyan). The black line is the median effective radius taken from K62 model fits to the observations}
  \label{fig:rhsmooth_Br_wobs}
\end{figure}

We combine our findings in Figure \ref{fig:rhsmooth_Bcompare_wobs} with the $\chi^2$ testing to create a possible anisotropy profile of M87 such that $\beta = 0$ for $R_{gc} \le 2 $ kpc and $\beta = 0.8$ for $R_{gc} \sim 10 $ kpc. We chose to simulate cluster populations with $\beta$ increasing proportional to $R_{gc}^{\frac{1}{4}}$ beyond 2 kpc. The effective radius distribution produced by this profile is plotted in Figure \ref{fig:rhsmooth_Br_wobs}. For convenience, we have also plotted the results of the $\beta=0$ simulation from Figure \ref{fig:rhsmooth_B0} in cyan. While this possible anisotropy profile yields a stronger agreement between theoretical and observational tidal radii, we view this profile as only a preliminary one and a broader range of anisotropy profiles should be explored. Nevertheless, our exploration of the range of $\beta$-models indicates that the imposition of a $\beta$-gradient is capable of matching the observed $r_h(R_{gc})$ trend seen so far.

Our findings that $\beta$ may increase with galactocentric distance are generally consistent with previous observational and theoretical results. However, having $\beta$ increase very quickly to a value of 0.8 at 10 kpc has not been found by previous studies. With a more gradual change in $\beta$, we would see that theoretical Jacobi radii are greater than observed K62 tidal radii in the mid-halo region. This can partially be attributed to clusters under-filling their tidal radius, as discussed in Section 4.1. In the inner regions of M87, where the tidal field is much stronger and globular clusters are expected to be tidally filled, we see a strong agreement between observationally determined and theoretical calculated cluster sizes.

\section{Conclusions and Future Work \label{sfive}}

We have measured accurate effective radii of approximately 2000 globular clusters within 10 kpc of the center of the giant elliptical galaxy M87. A theoretical cluster population was then simulated with the same observational characteristics of M87 and the tidal radius of each cluster was determined using the formalism of \citet{bertin08}.

The relationship between median cluster size and projected galactocentric distance was used to compare theoretical and observational tidal radii. To first order, it appears that the observational and theoretical cluster sizes are the same. Upon closer inspection, the theoretical and observational distributions are not in complete agreement, as tidal theory tends to overestimate cluster sizes in the outer regions of M87. Unfortunately, comparisons in the outer regions of M87 are constrained by the radial limit of our observations. In an upcoming HST Cycle 19 program, we will be able to add cluster size measurements extending to $R_{gc} \sim 80$ kpc, where stronger tests of the theory can be made.

One possible explanation for the discrepancy between theory and observations that we explore here is the effect of orbital anisotropy on the simulated distribution of cluster sizes. We compare simulations with different values of $\beta$ to observations, and conclude that M87 may have an anisotropy profile with $\beta \sim 0$ in the inner regions of the galaxy and $\beta > 0$ in the outer regions. This sort of profile is consistent with recent observational and theoretical results. We include a possible anisotropy profile which yields a stronger agreement between theoretical and observational tidal radii. 

Some issues about the cluster orbits and their internal dynamic evolution clearly remain to be investigated. Future work will include the use of Monte Carlo Markov Chain formalism to explore a broad range of anisotropy profiles to find the best possible match. Future N-body work will also explore the assumption that a cluster's tidal radius is imposed at perigalacticon. 

\section{Acknowledgements}

AS and WEH acknowledge financial support through research grants from the Natural Sciences and Engineering Research Council of Canada. We are indebted to John Blakeslee (HIA/NRC) for the generation of the composite HST images of M87.


\end{document}